\documentstyle[prl,aps,epsf]{revtex}
\begin{document}

\title {Unicellular algal growth: A biomechanical \\
approach to cell wall dynamics} 
\author {Royce Kam and Herbert Levine}
\address
{Department of Physics\\
University of California, San Diego\\
La Jolla, CA 92093-0402\\
}
\maketitle
\begin{abstract} 
We present a model for unicellular algal growth as 
motivated by several experiments implicating the
importance of calcium ions and ``loosening'' enzymes
in morphogenesis.  A  growing cell at
rest in a diffusive calcium solution is viewed as an
elastic shell on short timescales.  
For a given turgor pressure, we calculate
the stressed shapes of the wall elements whose
elastic properties are determined by Young's modulus
and the thickness of the wall.
The local enzyme concentration then determines the 
rate at which the unstressed shape of a wall 
element relaxes toward its stressed shape.
The local wall thickness is calculated from
the calcium-mediated addition of material
and thinning due to elongation.
We use this model to calculate growth rates for
small perturbations to a circular cell.  We find an
instability related to modulations of the wall thickness,
leading to growth rates which peak at a finite wave number.
\end{abstract} 
\pacs{}
In recent years, there has been increasing interest in models of 
unicellular growth.  Specific examples include dendritic branching
in neuronal growth~\cite{Hentschel,Albinet} and the broadening  
and branching of lobes in unicellular algal growth~\cite{PelceSun,PelcePocheau,Denet}.
The algal models have ranged from purely geometrical models (analogous to the
geometrical models of dendritic growth in solidification~\cite{Brower}) to one 
which combines the geometric formalism with a diffusive ``morphogen'' field.
At the same time, 
a wealth of experimental information regarding unicellular 
morphogenesis has been provided by studies of the 
larger species of the alga {\it Micrasterias}.
Morphogenesis in {\it Micrasterias} proceeds by a very 
well-ordered sequence of tip-splitting and lobe-broadening, 
culminating in elegant fan-shapes (Fig.1) which are reminiscent of
the patterns seen in unstable diffusion-controlled growth\cite{kkl}.  
This paper is devoted to the construction of a new model for
this process, a model built on the dynamics of the cell wall.

Although some of the details
remain elusive, a general picture of cell growth in algae has emerged from
experiment.
In order for the cell wall to elongate without thinning indefinitely,
vesicles containing the appropriate building materials are 
synthesized within the cytoplasm and then travel toward the cell 
periphery where the fusion of vesicles with the cell membrane
is thought to be mediated by calcium ions~\cite{Meindl}.
In particular, growing tips of {\it Micrasterias}
exhibit high concentrations of
membrane-associated calcium~\cite{Meindl,MeindlLancelle}, 
enabling them to fuse vesicles in
much larger quantities than other areas of the cell wall~\cite{Kiermayer}.
Meanwhile, there is evidence that calcium
concentration does not vary significantly within the body  of the cell 
itself~\cite{Holzinger}, perhaps implying that the relevant diffusive processes
occur {\it outside} the cell body even as they modulate the concentration at
the cell wall.

Experiments that halt growth by reducing turgor pressure 
demonstrate that elongation is (at least in part) a response
to the stresses in the cell wall~\cite{Kiermayer}.  In addition, it has long
been believed that a ``loosening factor'' must be present to allow the
fibers making up the wall to slip past each other during 
elongation (e.g. the protein ``expansin''\cite{Cosgrove}).
Since most elongation occurs at the cell tips 
~\cite{Meindl,Nishimura}, we might surmise that high
concentrations of calcium
also imply high concentrations of the ``loosening'' factor. 
These observations have led 
us to a simple model for cell wall dynamics in algae. 

Consider a cell growing in a solution of calcium ions.
We treat the cell wall as an elastic shell on short time scales,
whose slow plastic deformation is then catalyzed
by the concentration of a ``loosening'' factor.  We further
assume that the fusion of vesicles with the cell membrane (and subsequent
discharge of the vesicle contents into the cell wall) is also a
long time scale process.

We begin by discretizing the stressed shape into 
$N$ rods with lengths $l_i$ and orientations $\theta_i$.
Likewise, the unstressed shape will consist of $N$ rods with lengths
$l_i^\prime$ and orientations $\theta_i^\prime$ (and our convention hereafter will
be to associate all primed quantities with the unstressed shape).  The stressed
shape minimizes the energy

\begin{figure}
\epsfxsize=4.5in
\begin{center}
\mbox{\epsffile{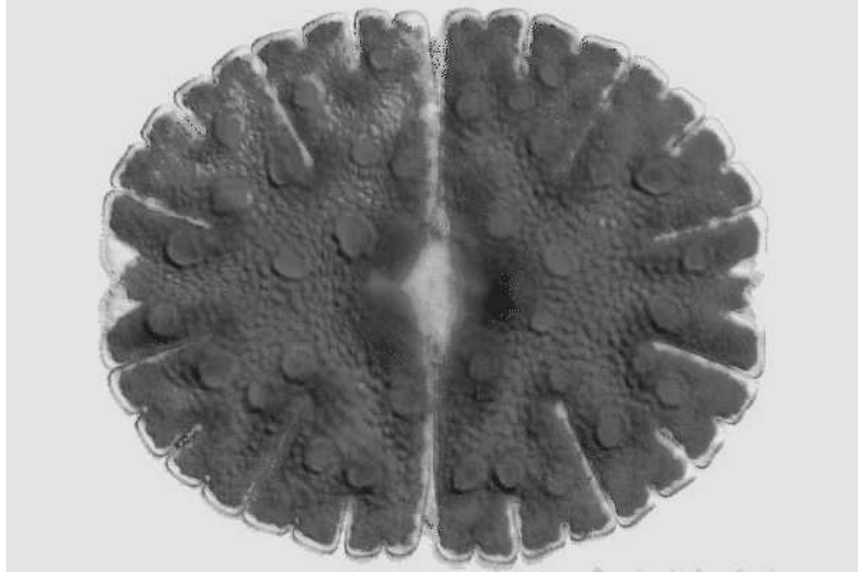}}
\end{center}
\caption{{\it Micrasterias denticulata}, diameter $\sim 200\mu m$. Picture provided
by U. Meindl.}
\end{figure}

\begin{eqnarray}
U = p\sum_{i=1}^{N}{l_i\sin{\theta_i}\sum_{j=1}^{i}{l_j\cos{\theta_j}}} +\sum_{i=1}^{N}{{1 \over 2}\alpha_i \left( l_i - l_i^\prime \right)^2} +\sum_{i=1}^{N}{{1 \over 2}\beta_i \left( \kappa_i - \kappa_i^\prime \right)^2}
\label{Utotal}
\end{eqnarray}
where the angle made by the intersection of two successive rods gives rise to the curvatures
$\kappa_i\approx \left( \theta_{i+1} - \theta_i \right)/  \frac{1}{2}\left( l_{i+1} + l_i \right)$. 
The sums are, respectively, the energy due to the pressure ($p\oint{xdy}$, where $p<0$ for an  outward pressure), the usual strain energy arising from Hooke's Law (where $\alpha_i\equiv Eh_i/l_i^\prime$ and $E$ is Young's modulus), and the pure bending energy of a rod (where $\beta_i\equiv EI_i l_i^\prime$
and the moment $I\equiv \int x^2 dx$ integrated over the rod's width
for an isotropic material\cite{Landau,2dRod}).
Note that if the cell has no known rotational symmetry, we must also constrain the
cell shape to be closed.

Once the stressed shape is determined, we can quasistatically calculate the
calcium concentration at the cell boundary.  As suggested by experiments, 
we assume that diffusion occurs {\it outside} the cell,
with the concentration $C(R_\infty)\equiv C_\infty$.
We therefore solve $\nabla^2u = 0$ in the region outside the cell, 
where $u\equiv(C-C_\infty)/C_\infty$ 
is the normalized concentration and $u(R_\infty)=0$.  
At the cell boundary itself, we specify the flux through the membrane with 
$\partial u/\partial n = j(u)/D$,
where {\bf n} is the {\it outward} normal to the cell surface, $j(u)$ is the {\it inward}
current, and $D$ is the diffusion constant.  

We then allow the unstressed shapes to relax toward the stressed configuration via
\begin{eqnarray}
\frac{dl_i^\prime}{dt~} = \Gamma(u) \left( l_i-l_i^\prime \right)
\label{ElongationRate}
\end{eqnarray}
\begin{eqnarray}
\frac{d\theta_i^\prime}{dt~} = \Gamma(u) \left( \theta_i-\theta_i^\prime \right)
\label{BendingRate}
\end{eqnarray}
where $\Gamma(u)$ is the characteristic relaxation rate as a function of the ``loosening'' factor
assuming, for simplicity, that it is the equivalent of the calcium concentration.
Finally, if we let $\chi\left( u \right)$ be the rate at which mass
(equivalently, the area $h_i l_i^\prime$, if the density is uniform)
is added to a given rod through the fusion of vesicles, then 
\begin{eqnarray}
\frac{dh_i}{dt~} =\frac{ \chi\left( u \right)}{l_i^\prime} - \frac{h_i}{l_i^\prime} \frac{dl_i^\prime}{dt~},
\label{ThinningRate}
\end{eqnarray}
which completes the specification of the model.

We can now use this model to calculate the growth rates 
of perturbations to the cell shape.
As the physics of the problem is invariant under reparametrizations, we can always
choose to parametrize the unstressed membrane by dividing it into $N$ rods of equal length
$l_i^\prime = a_0^\prime$. 
Then $\theta_i^\prime = \pi/2 + \phi_i + b_1^\prime\delta \sin{m\phi_i}$ with $\phi_i\equiv 2\pi\left( i-1 \right)/N$, specifies a perturbed circle with a curvature $\kappa^\prime\approx R_0^{\prime -1}+mb_1^\prime R_0^{\prime -1}\delta\cos{m\phi}$
where $R_0^\prime \equiv Na_0^\prime/2\pi$.
We specify the perturbed wall thickness as $h_i=h_0+h_1\delta\cos{m\phi_i}$.
We expect the stressed coordinates to have the form 
$l_i=a_0+a_1\delta\cos{m\phi_i}$
and $\theta_i=\pi/2 + \phi_i + b_1\delta \sin{m\phi_i}$.  [{\it Note}: Hereafter, 
we will refer to the pieces of a function $F$ expanded in $\delta$ by $F\equiv F_0 + F_1\delta\cos{m\phi}$.]

In the continuum limit, we expand the total energy as
$U  \approx  U_0 + \delta^2 U_2$, noting that the ${\cal O}(\delta)$ energy vanishes
upon integration with respect to $\phi$\cite{AreaNote}.  We solve the ${\cal O}\left( \delta^0 \right)$ problem 
by minimizing $U_0$ with respect to $a_0$, yielding a quartic equation for 
$a_0 \equiv 2\pi R_0/N$.
\begin{eqnarray}
R_0^4 \left( p + \frac{E h_0}{R_0^\prime} \right) - E h_0 R_0^3 + E I_0 R_0 - E I_0 R_0^\prime = 0
\end{eqnarray}
While we could obtain the {\it exact} solution for $a_0$ by taking the appropriate 
root, it is more enlightening to expand the solution to ${\cal O}\left( p \right)$, which 
gives 
 $R_0 \approx R_0^\prime - p R_0^{\prime^4}/(E h_0 R_0^{\prime^2} +E I_0)$.
we can solve for the perturbations by minimizing $U_2$ with respect to $a_1$ and $b_1$
(a long but straightforward calculation),
yielding an answer in terms of $a_0$~\cite{m1Case}.  
\begin{eqnarray}
a_1 =\left( a_0 -a_0^\prime \right)\frac{ p R_0^8 h_1 - \left( m^2-1 \right)  E I_0 R_0 R_0^\prime \left\{  m I_0 R_0  b_1^\prime+ \left[ R_0^3 +\frac{dI~}{dh_0}\left( R_0 -R_0^\prime \right) \right] h_1\right\}}{-p h_0 R_0^8 + \left( m^2-1 \right) E I_0 R_0^\prime \left[h_0 R_0^4- I_0 \left( R_0 -R_0^\prime \right)^2  \right]}
\label{a1Sol}
\end{eqnarray}
\begin{eqnarray}
b_1 = \frac{m \left( m^2-1 \right) E I_0 R_0 \left[h_0 R_0^4 -I_0 R_0^\prime\left( R_0 -R_0^\prime \right)  \right] b_1^\prime 
+ \left( R_0 -R_0^\prime \right) \left\{ p  R_0^7 -\left( m^2-1 \right)  {\cal B}   \right\} h_1}{-m p h_0 R_0^8 + m \left( m^2-1 \right) E I_0 R_0^\prime \left[ h_0 R_0^4-I_0 \left( R_0 -R_0^\prime \right)^2  \right]}
\label{b1Sol}
\end{eqnarray}
where ${\cal B} \equiv E R_0^3( I_0 R_0^\prime -h_0\frac{dI~}{dh_0} R_0)-E I_0( R_0 -R_0^\prime)( R_0^3 -R_0^\prime \frac{dI~}{dh_0} ) $, $I_0 \equiv I(h_0)$, and
$dI/dh_0$ is the derivative evaluated at $h=h_0$.
Observe that if the bending energy is made to vanish 
for a cell with uniform thickness (i.e. $I_0=dI/dh_0=h_1=0$)
we find that the stressed shape is a perfect circle ($a_1=b_1=0$).
This agrees with the well-known ``membrane'' result that the 
tension $T=|p|/\kappa$, implying that curvature
variations in a pliable membrane require external support\cite{Calladine}.

We must now calculate the concentration on the stressed shape, which by
direct integration of $dx_i = l_i \cos{\theta_i}$ and $dy_i = l_i\sin{\theta_i}$ is seen
to have the radial perturbation  $R_1 =R_0 \left(m b_1-a_1/a_0 \right)/(m^2-1)$ for $m>1$.~\cite{RadiusNote} 
The unperturbed solution is easily found to be
$u_0\left( r \right) = R_0 D^{-1} j_0 \ln(r/R_\infty)$ where $j_0=j(u_0(R_0))$.
Meanwhile, the perturbed solution must have the form
$u\left( r,\phi \right) \approx u_0\left( r \right) +c r^{-m} \delta\cos{m\phi}$.
Applying the flux boundary condition, 
we find the concentration at the cell wall, $u(R)\equiv u_0 + u_1\delta\cos{m\phi}$, to be \cite{ConcentrationNote}
\begin{eqnarray}
u\left( R \right) \approx u_0\left( R_0 \right) + \frac{j_0 \left( m-1 \right)}{m D + R_0\frac{\partial j}{\partial u_0}} R_1 \delta\cos{m\phi},
\end{eqnarray}
where $\partial j / \partial u_0$ (and like expressions will) refer(s) to the derivative evaluated at $u=u_0(R_0)$.

The stressed and unstressed shapes, along with the concentration just calculated,
allow us to calculate $l_i^\prime\left( t+dt \right)$ and $\theta_i^\prime\left( t+dt \right)$
through equations (\ref{ElongationRate}) and (\ref{BendingRate}).
Reparameterizing the solution at $t+dt$
so that we again have equal length rods~\cite{Reparametrize} yields
\begin{eqnarray}
\frac{d b_1^\prime}{dt~} = \Gamma_0 \left( b_1 -b_1^\prime \right) - \frac{\Gamma_0 a_1}{m a_0^\prime} +\frac{\partial \Gamma~}{\partial u_0} \frac{\xi_0 u_1}{m}.
\label{TangentRateCalc}
\end{eqnarray}
where $\Gamma_0\equiv \Gamma\left( u_0\left( R_0 \right) \right)$.
The rate of wall-thickening can be calculated directly from (\ref{ThinningRate}) as
\begin{eqnarray}
\frac{dh}{dt} = \frac{\chi_0}{a_0^\prime} -h_0\Gamma_0\xi_0 +\left\{ u_1 \left( \frac{1}{a_0^\prime} \frac{\partial \chi~}{\partial u_0} +h_0\frac{\partial \Gamma~}{\partial u_0}\xi_0 \right) +\Gamma_0 \left( h_1 \xi_0 +h_0\frac{a_1}{a_0^\prime} \right)   \right\}\delta\cos{m\phi}
\label{ThinningCalc}
\end{eqnarray}
where $\chi_0\equiv \chi\left( u_0\left( R_0 \right) \right)$ and 
$\xi_0\equiv (a_0-a_0^\prime)/a_0^\prime$.  
Since the average wall thickness is observed to be relatively
constant during the growth of the cell\cite{Meindl}, we must choose $\chi_0 = a_0^\prime h_0\Gamma_0\xi_0$.  We have tested these analytic results against simulations of the instantaneous
rates and found them to be in agreement, with the error converging as $1/N^2.$

Though these equations are somewhat complicated, we can still use them to 
examine qualitative behavior in several simple situations.
For definiteness, we assume that both the relaxation rate
and the vesicle fusion rate increase with higher concentrations 
(i.e. $\partial \Gamma/\partial u_0$, $\partial \chi/\partial u_0>0$).  The bending moment is
taken as $I\propto h^n$ where $n>1$ and $h\ll R$\cite{2dRod}.  Also, in accordance with
experimental observations\cite{Meindl}, we assume $j\left( u \right)$ causes an
inward current at the cell tips (e.g. $j=u-c_u$ with $c_u$ a constant).  For later reference, note that
the stressed curvature is $\kappa\approx R_0^{-1} + R_0^{-1}\left( m b_1 -a_1/a_0 \right)\delta\cos{m\phi}$.

First, consider a perfectly circular cell that develops develops
a slight thickening of the cell wall at $\phi=0$ (i.e. $b_1^\prime=0$ and $h_1>0$).
For small pressures we find that (to lowest order) 
$b_1 =-\xi_0 h_1 ( I_0 -h_0 \frac{dI}{dh_0} )/m I_0>0$ and 
$a_1 = -h_1(a_0-a_0^\prime)/h_0<0$.
This implies that $\phi=0$ is a ``tip'' ($\kappa_1 >0$), which is
also a minimum of strain ($a_1 <0$).
This agrees with the experimental observation that wall stresses
are minimized at cell tips\cite{Hejnowicz}.
Since $u_1>0$ at tips, $\Gamma_0 h_0 a_1/a_0$  is the only stabilizing term in (\ref{ThinningCalc}).  Therefore, if $h_0 |a_1|/a_0$ is not too large, the thickness variations
will grow, providing positive feedback to the instability which sprouted this tip.

Another interesting case is that of a perturbed cell shape which has a constant
thickness (i.e. $b_1^\prime>0$ and $h_1=0$).  To lowest order in pressure, we 
find that 
$(b_1 -b_1^\prime) \sim p b_1^\prime R_0^{\prime^3}/E h_0^{n-1}\left( m^2-1 \right) <0$,
$a_1\sim -\left( a_0-a_0^\prime \right)m b_1^\prime h_0^{n-1}/R_0^2<0$,
and $\kappa_1 \sim m b_1^\prime$.  
Again, we find that the minimum strain occurs at the ``tip''.
Looking at equation (\ref{TangentRateCalc}), we see that 
$-\Gamma_0 a_1/a_0$ is destabilizing, while $m\Gamma_0 (b_1-b_1^\prime)$
is stabilizing.  But since $h\ll R_0$, we expect the net effect to stabilize $b_1^\prime$.
Apparently, without modulations in the cell wall thickness, tips will be smoothed out.
It is worth noting that this effect may explain the ``lobe-broadening'' observed in later stages of
tip growth, though only  full numerical simulations would demonstrate this.

A trivial yet interesting implication of equation (\ref{ThinningCalc})
is the ability to reproduce the observed patterns of deposition 
of wall material when the turgor pressure
is reduced~\cite{Kiermayer}.  Setting $p=0$ implies that $\xi_0=a_1=0$, eliminating all stabilizing influences in the thickness and implies 
$dh_1/dt = u_1\frac{d\chi}{du_0}/a_0^\prime$.  That is,
the thickness variations will simply follow the variations in concentration, allowing
large amounts of material to collect at tips.  

Finally, we can use equations (\ref{TangentRateCalc}) and (\ref{ThinningCalc})
to obtain a pair of equations of the form
$db_1^\prime/dt = c_{11} b_1^\prime + c_{12}h_1$ and
$dh_1/dt = c_{21} b_1^\prime + c_{22} h_1$.
By assuming the coefficients vary slowly and   
letting $b_1^\prime, h_1 \sim exp(\lambda t)$, we can obtain an
expression for the quasi-static growth rates.  
While it is unprofitable to write the expression here,
we nonetheless plot the result for a typical set of
parameters in Figure 2.  We find that small-scale disturbances
are damped out, with the growth rate peaking at a finite wave number
determined by the physical parameters of the model.  
In addition, we note that the $m=1$ perturbation 
is indeed a ``zero-mode.''
\\
\begin{figure}
\epsfxsize=2.75in
\begin{center}
\mbox{\epsffile[98 225 512 520]{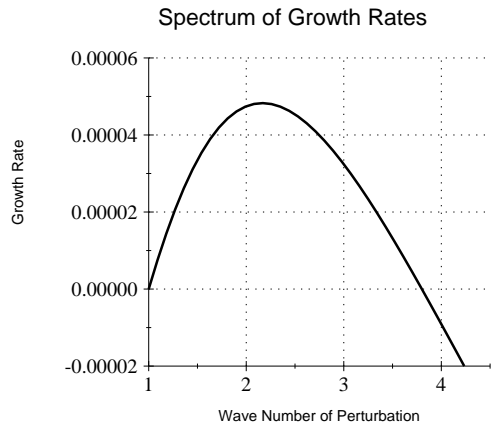}}
\end{center}
\caption{Spectrum of eigenvalues in the quasistatic approximation.  Growth rates ($\lambda$)
are plotted against wave number ($m$) for
$p = -0.001$, $E =20.0$, $h_0 =0.4$, $R_0^\prime=10.0$, $R_\infty=10^6$, $D=1$, $j = (u +0.15)$, $\Gamma = u +0.15$, $\chi = 0.12 \left( u +0.15 \right)$.}
\end{figure}

It is interesting to see that this relatively simple model can exhibit
such rich behavior and reproduce several experimentally observed effects.
This model has explicity assumed that reshaping of the cell wall is a relaxational process, wherein turgor pressure deforms the wall while enzymes allow the 
wall elements to slowly assume these stressed forms as their permanent forms.
Our model provides a way to understand these long-standing biological assumptions
analytically.  It also forms the physical basis for numerical simulations 
that may provide further insight into the mechanisms of morphogenesis 
at later stages of growth.

This work has been supported in part by NSF Grant No. DMR94-15460, and by the San Diego chapter of the ARCS Foundation.


\begin{thebibliography}{99}

\bibitem{Hentschel} Hentschel, H.G.E., Fine, A., {\em Phys. Rev. Lett.} {\bf 73}, 3592-3595 (1994).

\bibitem{Albinet} Albinet, G., Pelce, P., {\em Europhysics Letters} {\bf 33}, 569-574  (1996).

\bibitem{PelceSun} Pelce, P., Sun, J., {\em J. Theor. Biol.} {\bf 160}, 375-386 (1993).

\bibitem{PelcePocheau} Pelce, P., Pocheau, A., {\em J. Theor. Biol.} {\bf 156}, 197-214 (1992).

\bibitem{Denet} Denet, B., {\em Phys. Rev. E} {\bf 53}, 986-992 (1996).

\bibitem{Brower} Brower, R.C., Kessler, D.A., Koplik, J., Levine H., {\em Phys. Rev. Lett.} {\bf 51}, 309-312 (1983).

\bibitem{kkl} D. A. Kessler, J. Koplik and  H. Levine, {\em Advances
in Physics} {\bf 37}, 255-339 (1988).

\bibitem{Meindl} Meindl, U.,{\em Microbiological Reviews} {\bf 57}, 415-433 (1993).

\bibitem{MeindlLancelle} Meindl, U., Lancelle, S., Hepler, P., {\em Protoplasma} {\bf 170}, 104-114 (1992).

\bibitem{Kiermayer} Kiermayer, O. ,{\em Cytomorphogenesis in Plants}, ed. O. Kiermayer,
Springer-Verlag, Wien New York (1993), 147-189.

\bibitem{Cosgrove} Cosgrove, D.,{\em BioEssays} {\bf 18}, 533-540 (1996).

\bibitem{Nishimura} Nishimura, M., Ueda, K.,{\em J. Cell Science} {\bf  31}, 225-231 (1978).

\bibitem{Landau} Landau, L.D. and Lifshitz, E.M.,  {\em Theory of Elasticity}, Pergamon Press
, London, 1959.  See pages 75-79 for a derivation of the bending energy of a rod.

\bibitem{2dRod}  For our 2d rod, $I_i = h_i^3/12$.

\bibitem{AreaNote}  For $m>1$, 
$A_1=-\pi R_)^2\left( mb_1-a_1/a_0 \right)^2/2\left( m^2-1 \right)$
and $A_1 = -\pi R_0^2\left( b_1 -a_1/a_0 \right)^2/8$ for $m=1$.

\bibitem{Holzinger} Holzinger, A.,Callaham, D., Hepler, P., Meindl, U.,{\em European Journal of Cell Biology} {\bf 67}, 363-371 (1995).

\bibitem{AreaNote}  The strain and bending energies are straightforward.  The perturbation
to the area is  \\
$A_1=-\frac{1}{2}\pi R_0^2\left( mb_1-a_1/a_0 \right)^2/\left( m^2-1 \right)$ for $m>1$,
and $A_1 = -\frac{1}{8}\pi R_0^2\left( b_1 -a_1/a_0 \right)^2$ for $m=1$.

\bibitem{Calladine} Calladine,C.R.,{\em Theory of Shell Structures},Cambridge University Press,
 80-86 (1983).

\bibitem{Reparametrize}  Given $l = A_0 + A_1\delta\cos{m\phi}$ and $\theta=\pi/2 + \phi + B_1\delta\sin{m\phi}$, the equal-length parametrization yielding the same curvature
has $A_{0,new}= A_0$ and $B_{1,new}=  B_1 - A_1/m A_0$.

\bibitem{m1Case} For $m=1$, the total $y$-displacement $\propto \left( b_1 -a_1/a_0 \right)$.  To ensure a closed shape, we impose $b_1=a_1/a_0$.  Minimization yields $a_1=-h_1 \left( a_0-a_0^\prime \right)/h_0$, which
agrees with the limit $m\rightarrow 1$ of equation (\ref{a1Sol}).

\bibitem{RadiusNote} For $m=1$, explicit integration of $x\left( \phi \right)$ and $y \left(\phi  \right)$ shows that $R_1=\frac{1}{2}R_0 \left( b_1 + a_1/a_0\right)$.

\bibitem{ConcentrationNote}  Observe that $u_1=0$ for $m=1$, though 
technically speaking we have not required the perturbation to vanish at $r=R_\infty$.

\bibitem{Hejnowicz}Hejnowicz, Z., Heinemann, B., and Sievers, A.,{\em Z. Pflanzenphsiol. Bd.}{\bf 81}, 409-424 (1977).

\end{thebibliography}
\end{document}